\documentclass[10pt,conference]{IEEEtran}
\IEEEoverridecommandlockouts
\pdfoutput=1

\usepackage[letterpaper, left = 0.68in, right = 0.68in, top = 0.75in, bottom = 1.1in]{geometry}

\usepackage{cite}
\usepackage{amsmath,amssymb,amsfonts,stmaryrd}
\usepackage{algorithmic}
\usepackage{graphicx}
\usepackage{textcomp}
\usepackage{xcolor}
\usepackage{hyperref}
\usepackage[nolist]{acronym} 
\usepackage{calc}
\usepackage{tikz}
\usepackage{pgfplots}
\usepgfplotslibrary{fillbetween}
\usetikzlibrary{automata,positioning}
\usetikzlibrary{decorations.pathreplacing}
\usetikzlibrary{patterns, arrows}
\usepackage{standalone}

\definecolor{mittelblau}{RGB}{0, 126, 198}
\definecolor{rot}{RGB}{238, 28 35}
\definecolor{apfelgruen}{RGB}{140, 198, 62}
\definecolor{gelb}{RGB}{255, 229, 0}
\definecolor{orange}{RGB}{244, 111, 33}
\definecolor{pink}{RGB}{237, 0, 140}
\definecolor{lila}{RGB}{128, 10, 145}
\definecolor{hellgrau}{RGB}{224, 224, 224}
\definecolor{mittelgrau}{RGB}{128, 128, 128}
\definecolor{dunkelgrau}{RGB}{80,80,80}
\definecolor{anthrazit}{RGB}{19, 31, 31}
\definecolor{darkgreen}{RGB}{34,139,34}

\def\lw{1.0pt}

\newtheorem{thm}{Theorem}
\renewcommand{\IEEEQED}{\IEEEQEDopen}

\begin{document}

\title{Bounds for Joint Detection and Decoding \\on the Binary-Input AWGN Channel
}

\author{\IEEEauthorblockN{Simon Obermüller, Jannis Clausius, Marvin Rübenacke, and Stephan ten Brink}
	\IEEEauthorblockA{
		Institute of Telecommunications, Pfaffenwaldring 47, University of  Stuttgart, 70569 Stuttgart, Germany 
		\\\{obermueller,\;clausius,\;ruebenacke,\;tenbrink\}@inue.uni-stuttgart.de\\
	}
    \thanks{This work is supported by the German Federal Ministry of Education and Research (BMBF) within the project Open6GHub (grant no. 16KISK019).}
}

\maketitle
\begin{NoHyper}
\begin{acronym}
\acro{ML}{maximum likelihood}
\acro{BER}{bit error rate}
\acro{SNR}{signal-to-noise-ratio}
\acro{BPSK}{binary phase shift keying}
\acro{AWGN}{additive white Gaussian noise}
\acro{BI-AWGN}{binary-input additive white Gaussian noise}
\acro{LLR}{log-likelihood ratio}
\acro{MAP}{maximum a posteriori}
\acro{ML}{maximum likelihood}
\acro{BEC}{Binary Erasure Channel}
\acro{BSC}{Binary Symmetric Channel}
\acro{3GPP}{3rd Generation Partnership Project }
\acro{eMBB}{enhanced Mobile Broadband}
\acro{URLLC}{ultra-reliable low-latency communications}
\acro{mMTC}{massive machine-type communications}
\acro{IoT}{internet of things}
\acro{CCDF}{complementary cumulative distribution function}
\acro{QAM}{quadrature amplitude modulation}
\acro{JDD}{joint detection and decoding}
\acro{HyPED}{hybrid preamble and energy detection}
\acro{DAD}{decoder-aided detection}
\acro{PBD}{prior-based detection}
\acro{UB}{union bound}
\acro{DT}{dependency testing}
\acro{i.i.d.}{independent and identically distributed}
\end{acronym}

\begin{abstract}
For asynchronous transmission of short blocks, preambles for packet detection contribute a non-negligible overhead. To reduce the required preamble length, \ac{JDD} techniques have been proposed that additionally utilize the payload part of the packet for detection.
In this paper, we analyze two instances of \ac{JDD}, namely \ac{HyPED} and \ac{DAD}.
While \ac{HyPED} combines the preamble with energy detection for the payload, \ac{DAD}  also uses the output of a channel decoder.
For these systems, we propose novel achievability and converse bounds for the rates over the \ac{BI-AWGN} channel.
Moreover, we derive a general bound on the required blocklength for \ac{JDD}.
Both the theoretical bound and the simulation of practical codebooks show that the rate of \ac{DAD} quickly approaches that of synchronous transmission.
\end{abstract}

\acresetall %

\section{Introduction}

The rapid evolution of wireless communication systems has given rise to a wide range of applications, each with different performance requirements. In particular, the \ac{IoT}, \ac{mMTC}, and \ac{URLLC} represent three pillars of next-generation networks characterized by the need for high efficiency, scalability, and minimal latency \cite{mahmood2021machinetype}.
Each of these systems is characterized by asynchronous channel access, requiring means for the receiver to detect and synchronize to a transmitted packet.
Traditionally, this is done using a dedicated, deterministic preamble \cite{massey1972synchronization} designed for good synchronization performance.
Assuming a synchronized packet, a conventional channel code is used to protect the message content from errors.  %
However, it has been shown that packet detection only using dedicated preambles is suboptimal \cite{wang2011errorexponents}.
For \ac{IoT} applications, the short blocklength regime is of special interest \cite{vaezi2022cellular}. 
Here, the overhead introduced by the preamble is non-negligible.
\Ac{JDD} additionally uses the payload portion of the packet for the detection task, allowing the preamble size to be reduced or eliminated, improving battery life and reducing channel utilization and latency \cite{lancho2021jdd}. The frame structure for both approaches is depicted in Fig.~\ref{fig:structure}.

Recent works mainly focus on designing practical \ac{JDD} schemes. 
A classical and a machine-learning based approach are considered in \cite{saied2022ccsk} and \cite{dorner2022learning}, respectively.
Furthermore, schemes for coexisting synchronous and asynchronous communications \cite{marata2023detectionanddecoding} and multi-user random access \cite{Zhang2024Jointestimation} have been proposed.
However, the theoretical foundations and limitations of \ac{JDD} are not yet fully established.

The modern model for asynchronous transmissions is first presented and theoretically analyzed in \cite{tchamkerten2009asynchronism}.
A particular instance of the problem is \textit{slotted asynchronous} transmission, i.e., the receiver is only required to detect whether a packet is present or not (and then decode the packet), rather than to synchronize in time.
For this problem, \cite{wang2011errorexponents} introduces and analyzes the error exponents of heuristic \ac{HyPED}, a detection scheme that combines correlation-based preamble detection with energy detection on the message.
In \cite{weinberger2014jdd}, the authors establish the optimal detection rule for the slotted asynchronous communication scenario and their error exponents. However, the optimal detection involves computing the likelihood of all possible codewords, which is generally infeasible.
Therefore, the same authors propose a simplified detection rule, namely \ac{DAD}, that only requires to compute the likelihood of the codeword estimate of a (potentially suboptimal) channel decoder \cite{weinberger2017channeldetection}.
Recently, finite blocklength bounds for \ac{JDD} on the information rate were derived \cite{lancho2021jdd}.
To facilitate numerical evaluation, a detection rule was proposed based on skewed priors of payload symbols, which are assumed to be statistical independent.

\begin{figure}[t]
\centering\resizebox{0.8\linewidth}{!}{\begin{tikzpicture}

    \draw[line width=1pt, color=orange, fill=orange!30, text=black] (0,0) rectangle (3,-0.7) node[pos=.5] {Preamble};
    \draw[line width=1pt, color=rot, fill=rot!30, text=black] (3,0) rectangle (8,-0.7) node[pos=.5] {Codeword};
    \draw[line width=1pt, color=apfelgruen, fill=apfelgruen!30, text=black] (2,-2) rectangle (8,-1.3) node[pos=.5] {Codeword};
    
    \path[draw,decorate,decoration={brace,amplitude=5pt}] (8,-2.2) -- (2,-2.2)
    node[midway,below=5pt]{Joint Detection and Decoding};
    
    \path[draw,decorate,decoration={brace,amplitude=5pt}] (0,0.2) -- (3,0.2) 
    node[midway,above=5pt]{Detection\strut};
    \path[draw,decorate,decoration={brace,amplitude=5pt}] (3,0.2) -- (8,0.2) 
    node[midway,above=5pt]{Decoding};
    
    \draw[latex'-latex',very thick] (0,-1.65) -- (2,-1.65)
    node[midway,below=2pt,align=center]{saved\\channel uses};
    
    \draw[dashed] (0,-.7) -- (0,-2);
    \draw[dashed] (2,-.7) -- (2,-1.3);

\end{tikzpicture}}
\caption{Frame structure for conventional, separate detection and decoding (top) and \ac{JDD} (bottom).}
\label{fig:structure}
\end{figure}
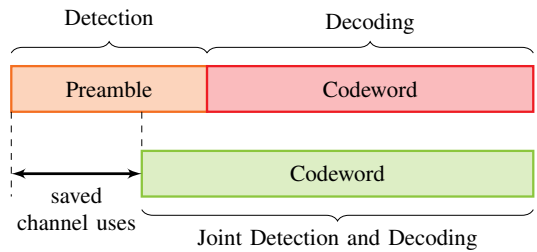

In this paper, we present new bounds for practical \ac{JDD} schemes over the \ac{BI-AWGN} channel. Our contributions are summarized as follows:
\begin{itemize}
    \item We derive the optimal decision rule for \ac{HyPED} and compute the corresponding achievability and converse bounds on the information rate.
    \item We establish a general converse bound for the required blocklength in the \ac{JDD} problem.
    \item For \ac{DAD}, we present an achievability bound on the information rate and show that practical coding schemes achieving this bound.%
\end{itemize}
The bounds show that both \ac{DAD} and \ac{HyPED} can outperform the scheme proposed in \cite{lancho2021jdd}, where only statistical properties of the codebook are utilized.

\paragraph*{Notation}
Scalars and vectors are given by normal font $x$ and bold lower case letters $\mathbf{x}$, respectively. The Euclidean norm is denoted by $\Vert \cdot  \Vert$. Random variables are capitalized, e.g., $X$ or $\mathbf{X}$.
The normal distribution with mean $\mu$ and variance $\sigma^2$ is denoted by $\mathcal{N}(\mu,\sigma^2)$ .
The corresponding complementary cumulative distribution is $Q(x)$. 
$\operatorname{Pr}\{x\}$ denotes the probability of the event $x$.
Finally, $\operatorname{p}(x)$ is the probability density function of the realization $x$ of the random variable $X$.

\section{Preliminaries}

\subsection{System Model and Performance Metrics}
We consider a slotted asynchronous system divided into time slots of length $n$. 
For every slot, the transmitter decides its state; either it is active and $m \in \mathcal{M} \triangleq \left\{1,...,M\right\}$, or it is idle and $m = \O$.
When being active, a message $m$ is uniformly drawn from $\mathcal{M}$ and mapped to a codeword $\mathbf{c} \in \mathbb{F}_2^n$.
Subsequently, $\mathbf{c}$ is modulated with \ac{BPSK} to $\mathbf{x}_m$.
Thus, the information rate is $R=\log_2(M)/n$. 
If the transmitter is idle ($m=\O$), the channel input is the all-zero vector $\mathbf{x}_{\O}=\mathbf{0}$ denoted as an empty slot.
The observations at the receiver after addition of the \acs{AWGN} noise $z \sim \mathcal{N}\left(0,\sigma^2 \right)$ with \ac{SNR} $E_\mathrm{S}/N_0 =\frac{1}{2\sigma^2}$ is called $\mathbf{y}$.
The receiver estimates for each slot whether the transmitter has been idle, $\hat{m}=\O$, or active, $\hat{m}\in \mathcal{M}$ (detection).
If a transmission is detected, the receiver estimates the message $\hat{m} \in \mathcal{M}$ (decoding).

The performance of the system is measured according to the following error rates.
The false alarm rate $P_\mathrm{FA}$ and missed detection rate $P_\mathrm{MD}$ describe  the detection problem and are defined as
\begin{align*}
    P_\mathrm{FA} &\triangleq \sum_{m=1}^{M} \mathrm{Pr} \left\{  \hat{m}=m \mid  \mathbf{x}_{\O} \right\}, \\
    P_\mathrm{MD} &\triangleq \frac{1}{M}\sum_{m=1}^{M} \mathrm{Pr} \left\{ \hat{m}=\O \mid  \mathbf{x}_m \right\}.
\end{align*}
The false alarm rate is the probability that the receiver detects a transmission even though the slot was empty.
Similarly, the missed detection rate is the probability of a slot containing a transmission being mistaken for an empty slot averaged over all transmissions.
Note that $P_\mathrm{MD} $ and $P_\mathrm{FA} $ oppose each other.
If one decreases, the other increases, resulting in a trade-off.
The optimal $P_\mathrm{MD} $ for $P_\mathrm{FA}\leq \epsilon_\mathrm{FA}$, given the constraint $\epsilon_\mathrm{FA}$, is described by the Neyman-Pearson lemma \cite{neyman}

\begin{equation*}
    \min_{\text{s.t. }P_\mathrm{FA} \leq \epsilon_\mathrm{FA}  } P_\mathrm{MD}.
\end{equation*}
The resulting decision rule is the likelihood ratio test 
\begin{equation}\label{eq:np_l_test}
    \frac{ \overbrace{\textstyle\sum_{m'=1}^{M} \mathrm{p} \left( \mathbf{y} \mid \mathbf{x}_{m'} \right)}^{\text{codebook likelihood}}}
    { \mathrm{p} \left(  \mathbf{y} \mid \mathbf{x}_{\O} \right)} 
    \mathrel{\substack{\O\\\lessgtr\\\mathcal{M}}}  \gamma,
\end{equation}
where the threshold $\gamma$ is chosen such that $P_\mathrm{FA} \leq \epsilon_\mathrm{FA}$. %

For the decoding problem, given an active transmission, the codeword error rate $P_\mathrm{CW}$ is defined as the probability that the decoding rule selects the wrong codeword 
\begin{equation}\label{eq:pcw}
   P_\mathrm{CW} \triangleq \frac{1}{M} \sum_{m=1}^{M} \mathrm{Pr} \left\{  \hat{m}  \neq m \mid \mathbf{x}_m, \hat{m}\neq \O \right\}.
\end{equation}

Finally, the inclusive error rate $P_\mathrm{IE}$ measures jointly the detection and decoding problem.
It is defined as the average probability that the receiver either missed a detection or decoded the incorrect message
\begin{equation}\label{eq:pie}
    P_\mathrm{IE} \triangleq \frac{1}{M} \sum_{m=1}^{M} \mathrm{Pr} \left\{ \hat{m} \neq m \mid \mathbf{x}_m \right\}.
\end{equation}
Observe that in equation (\ref{eq:pie}), the case $\hat{m}=\O$ is included, while for equation (\ref{eq:pcw}), $\hat{m}\neq\O$.

\subsection{Decoder-aided Detection and Decoding}
We first consider the optimal solution of the \ac{JDD} problem, i.e., the decision rule such that no other decision rule leads to strictly lower error rates $P_\mathrm{FA}, P_\mathrm{MD}$, and $P_\mathrm{IE}$ simultaneously. 
In \cite[Sec. 3]{weinberger2014jdd} the optimal decision rule for the joint detector and decoder is given by the threshold test
\begin{equation}\label{eq:optimal_det_dec}
    \frac{ \gamma_a \sum_{m'=1}^{M} \mathrm{p} \left( \mathbf{y} \mid  \mathbf{x}_{m'} \right)+\max_{m'\in \mathcal{M}} \operatorname{p}(\mathbf{y} \mid \mathbf{x}_{m'})}{ \mathrm{p} \left( \mathbf{y} \mid  \mathbf{x}_{\O} \right)}  \mathrel{\substack{\O\\\lessgtr\\\hat{m}}}  \gamma_b,
\end{equation}
where $\hat{m}=\arg \max_{ \mathbf{y} \in \mathcal{M}} \operatorname{Pr}\{m' \mid \mathbf{x}_{m'}\}$ is the \ac{ML} codeword, and $\gamma_a$ and $\gamma_b$ are chosen according to the desired trade-off between $P_\mathrm{FA}$ and $P_\mathrm{MD}$, while simultaneously minimizing $P_\mathrm{IE}$.
That means, the difference to the optimal detection rule in equation (\ref{eq:np_l_test}) is the addition of the likelihood of the \ac{ML} codeword to the codebook likelihood.
In other words, minimizing $P_\mathrm{IE}$ means that we only detect a transmission if we can decode it with a sufficiently high probability.
Since $\operatorname{p}(\mathbf{y}\mid \mathbf{x}_{m'})$ is computed for every codeword in the codebook, we label it codebook-aided detection.
A simplified decision rule is \acf{DAD} \cite[Sec. 4B]{weinberger2017channeldetection}
\begin{align}
&
  \frac{ \operatorname{p}(\mathbf{y}\mid\mathbf{x}_{\hat{m}})
    }
    {
    \operatorname{p}(\mathbf{y}\mid\mathbf{x}_{\O})
    }=
   \frac{
    \max_{m'\in \mathcal{M}} \operatorname{p}(\mathbf{y}\mid\mathbf{x}_{m'})
    }
    {
    \operatorname{p}(\mathbf{y}\mid\mathbf{x}_{\O})
    }
    \mathrel{\substack{\O\\\lessgtr\\\hat{m}}}
    \gamma' \label{eq:dad_sub_optimal_JDD} \\
     \stackrel{\text{BI-AWGN}}{\iff} &
     \max_{m'\in \mathcal{M}}\mathbf{x}_{m'}^\mathrm{T}\mathbf{y}
    \mathrel{\substack{\O\\\lessgtr\\\hat{m}}} \gamma. \nonumber %
\end{align}
Note that for $\hat{m}=m\in\mathcal{M}$ or $m=\O$ (independent of $\hat{m}$), the correlation follows a Gaussian distribution according to
\begin{equation}\label{eq:dad_lrt_normal}
    \ln\frac{
    \operatorname{p}(\mathbf{y}\mid\mathbf{x}_{\hat{m}})
    }
    {
    \operatorname{p}(\mathbf{y}\mid\mathbf{x}_{\O})
    } ~\propto~
    \mathbf{x}_{\hat{m}}^\mathrm{T}\mathbf{y} \sim 
    \begin{cases}
    \mathcal{N}(n,n\sigma^2), & \hat{m}=m \in \mathcal{M}\\
    \mathcal{N}(0,n\sigma^2), & m=\O.
\end{cases}
\end{equation} 
The distribution for all other  $\mathbf{x}_{\hat{m}}^\mathrm{T}\mathbf{y},~\hat{m}\neq m \in \mathcal{M}$ is dependent on the codebook.
Equation (\ref{eq:dad_sub_optimal_JDD}) is equal to equation (\ref{eq:optimal_det_dec}) for $\gamma_a=0$ or approximates it for high \acp{SNR}, where the codebook likelihood is $\sum_{m'=1}^{M} \mathrm{p} \left( \mathbf{y}  \mid  \mathbf{x}_{m'}\right) \approx  \operatorname{p}(\mathbf{y} \mid \mathbf{x}_{\hat{m}})$.
Using the same approximation in equation (\ref{eq:np_l_test}) also leads to equation (\ref{eq:dad_sub_optimal_JDD}).
Note, in cases where finding the \ac{ML} codeword is not feasible, a sub-optimal decision $\hat{m}'$ can be used by approximating the numerator  $ \operatorname{p}(\mathbf{y}\mid\mathbf{x}_{\hat{m}}) \approx\operatorname{p}(\mathbf{y}\mid \mathbf{x}_{\hat{m}'})$.
\Ac{DAD} simplifies the decision rule but not necessarily decreases the complexity, as some decoding algorithms output the codebook likelihood as byproduct, e.g., naive \ac{ML} decoding (that involves enumeration over the codebook anyways), or list-decoding of polar codes \cite{yuan24generalizedpolardecoding}.

\subsection{Heuristic Hybrid Preamble and Energy Detection}
In \cite[Sec. 4B]{wang2011errorexponents}, it is shown that detection based on both a known preamble and the energy of the unknown codeword outperforms detection solely on the preamble in terms of error exponent on an \ac{AWGN} channel. 
For this, we split $\mathbf{x}$ and $\mathbf{y}$ into a preamble part  $\mathbf{x}_\mathrm{p}$ and $\mathbf{y}_\mathrm{p}$ of length $n_\mathrm{p}$ and a codeword part $\mathbf{x}_\mathrm{c}$ and $\mathbf{y}_\mathrm{c}$ of length $n_\mathrm{c}=n-n_\mathrm{p}$.
The underlying assumption for energy detection are independent symbols $\operatorname{Pr}\left\{ \mathbf{x}\right\} = \prod_i \operatorname{Pr}\left\{ x_i\right\}$.
The heuristic \acf{HyPED} detection rule is then
\begin{equation}\label{eq:wang_heuristic_jdd_peramble_energy}
    \gamma_a\mathbf{y}_\mathrm{p}^\mathrm{T} \mathbf{x}_\mathrm{p}
    + \Vert \mathbf{y}_\mathrm{c} \Vert
    \mathrel{\substack{\O\\\lessgtr\\\hat{m}}} \gamma_b,
\end{equation}
where $\gamma_a$ and $\gamma_b$ are constants chosen according to the $P_\mathrm{FA}$ and $P_\mathrm{MD}$ requirements.
In section \ref{ssec:hyped}, we will generalize this heuristic and derive the optimal detection rule.

\subsection{Distribution-based Detection Bounds for Finite Length}
While the previous works on slotted asynchronous transmissions evaluated the error exponent, the authors of \cite{lancho2021jdd} show bounds on the rate in the finite blocklength regime. 
Similar to the  $\beta \beta$ bounds from \cite{yang2018beta}\cite{polyanskiy2013empirical}, converse and achievability bounds are derived for \ac{JDD} and preamble-based detection. 
Here, the input distribution is chosen as \ac{i.i.d.} $\operatorname{Pr}\{\mathbf{x}\}=\prod_i \operatorname{Pr}\{x_i\}$.
To facilitate detection, the individual priors are skewed $0<\operatorname{Pr}\{x_i=+1\}\leq 0.5$ which implies a reduced information rate.
Note that, in contrast, \ac{DAD} facilitates the detection by lowering the code rate and with it the information rate.
The resulting bounds for the \ac{BI-AWGN} channel are, thus, called  \ac{PBD} bounds.

For the case of genie-aided detection (i.e., $P_\mathrm{FA}=0$ and $P_\mathrm{MD}=0$), we evaluate the same bounds as in \cite{lancho2021jdd}.
These are the meta-converse bound \cite[Th. 15]{polyanski2010bounds}\cite{vazquez2018saddlepoint} and the \ac{DT} achievability bound\cite[Th. 17]{polyanski2010bounds}.

\section{Joint Detection and Decoding Schemes and their Bounds}

In this section, we analyze detection schemes that can be used in \ac{JDD} systems and derive short-blocklength bounds for the \ac{BI-AWGN} channel.
\subsection{Converse Bound on the Required Blocklength}
\begin{thm}\label{thm:converse_blocklength}
    For a system with noise variance $\sigma^2$ and required error rates $P_\mathrm{FA} \leq \epsilon_\mathrm{FA}$ and $P_\mathrm{MD} \leq \epsilon_\mathrm{MD}$, the blocklength $n$ is lower bounded by 
    \begin{equation}
        n \geq \sigma^2 \,  \cdot \left( Q^{-1} (\epsilon_\mathrm{FA})- Q^{-1}(1-\epsilon_\mathrm{MD}) \right)^2.
    \label{eq:general_converse_blocklength}
    \end{equation}
\end{thm}

\begin{IEEEproof}
Assume a system with a genie-aided receiver, that knows the message that will be transmitted the next time the transmitter is active. 
Then, the optimal detector from equation (\ref{eq:np_l_test}) simplifies to a likelihood ratio test
\begin{equation}\label{eq:conv_bound_llr_test}
    \frac{
    \operatorname{p}(\mathbf{y}\mid\mathbf{x}_m)
    }
    {
    \operatorname{p}(\mathbf{y}\mid\mathbf{x}_{\O})
    } 
    \mathrel{\substack{\O\\\lessgtr\\m}}
    \tilde{\gamma},
\end{equation}
where $\tilde{\gamma} \in \mathbb{R}$ is the decision threshold. 
Note that, even though the system models the detection and decoding problem, the genie removes the decoding part of the problem.
Thus, in order to achieve the optimal $P_\mathrm{FA}$ and $P_\mathrm{MD}$ trade-off, equation equation (\ref{eq:np_l_test}) is used.

Consequently, the false alarm and missed detection rate can be calculated as
\begin{align*}
    P_\mathrm{FA} &\stackrel{(\ref{eq:conv_bound_llr_test})}{=} \operatorname{Pr} \left\{ \frac{
    \operatorname{p}(\mathbf{y} \mid \mathbf{x}_m)
    }
    {
    \operatorname{p}(\mathbf{y}\mid\mathbf{x}_{\O})
    }  > \tilde{\gamma} \;\middle |\; \mathbf{x}_{\O} \right\} 
     \stackrel{(\ref{eq:dad_lrt_normal})}{=} Q \left( \frac{\gamma}{\sqrt{n \sigma^2}}\right), \\
    P_\mathrm{MD} &\stackrel{(\ref{eq:conv_bound_llr_test})}{=}  \operatorname{Pr} \left\{ \frac{
    \operatorname{p}(\mathbf{y}\mid\mathbf{x}_m)
    }
    {
    \operatorname{p}(\mathbf{y}\mid\mathbf{x}_{\O})
    }  \leq \tilde{\gamma} \;\middle |\; \mathbf{x}_{m} \right\} 
    \stackrel{(\ref{eq:dad_lrt_normal})}{=} 1 - Q \left( \frac{\gamma -n}{\sqrt{n \sigma^2}} \right).
\end{align*}
By inserting $P_\mathrm{FA} = \epsilon_\mathrm{FA}$ and $P_\mathrm{MD} = \epsilon_\mathrm{MD}$, this can be rearranged to obtain equation (\ref{eq:general_converse_blocklength}).
\end{IEEEproof}

This bound on the blocklength can also be understood as the minimum length for which the detection problem with respect to the requirements $\epsilon_\mathrm{FA}$ and $\epsilon_\mathrm{MD}$ can be solved. As no data is being sent, the receiver is either idle or transmits a fixed sequence which can be optimally detected by the receiver via a matched filter, i.e., using correlation.
\subsection{Achievability Bound for Decoder-aided Detection}
When the detector applies the \ac{DAD} scheme, the following achievability bound holds.

\begin{thm}\label{thm:dad_ach_bound}
    For a system detecting according to equation (\ref{eq:dad_sub_optimal_JDD}) with blocklength $n$ and noise power $\sigma^2$, there exists a code of size $M$ with
    \begin{align*}
        P_\mathrm{FA} &\leq M \cdot  Q \left( \frac{\gamma}{\sqrt{n\sigma^2}} \right),\\
        P_\mathrm{MD} &\leq 1- Q \left( \frac{\gamma-n}{\sqrt{n\sigma^2}} \right),\\
        P_\mathrm{IE} &\leq P_\mathrm{MD} + P_\mathrm{CW},
    \end{align*}
    where $\gamma \in \mathbb{R}$ is a fixed decision threshold and $P_\mathrm{CW}$ is bounded by an achievable error rate for synchronous transmission.
    
\end{thm}

\begin{IEEEproof}
We prove the achievability bound by upper bounding $P_\mathrm{FA}$, $P_\mathrm{MD}$ and $P_\mathrm{IE}$. Applying the \ac{UB} to $P_\mathrm{FA}$ results in 
\begin{align}
P_\mathrm{FA} &\stackrel{(\ref{eq:dad_sub_optimal_JDD})}{=} \mathrm{Pr} \left\{ 
\frac{ \operatorname{p}(\mathbf{y}\mid\mathbf{x}_{\hat{m}})
    }
    {
    \operatorname{p}(\mathbf{y}\mid\mathbf{x}_{\O})
    } 
\geq \gamma \;\middle |\; \mathbf{x}_{\O} \right\} \nonumber \\
&= \mathrm{Pr} \left\{ \exists \, m' \in \mathcal{M}: \frac{
    \operatorname{p}(\mathbf{y}\mid\mathbf{x}_{m'})
    }
    {
    \operatorname{p}(\mathbf{y}\mid\mathbf{x}_{\O})
    }  \geq \gamma \;\middle |\; \mathbf{x}_{\O} \right\} \nonumber \\
& \stackrel{\mathrm{UB}}{\leq} \sum_{m'=1}^{M} \mathrm{Pr} \left\{ \frac{
    \operatorname{p}(\mathbf{y}\mid\mathbf{x}_{m'})
    }
    {
    \operatorname{p}(\mathbf{y}\mid\mathbf{x}_{\O})
    }  \geq \gamma \;\middle |\; \mathbf{x}_{\O} \right\} \nonumber 
    \\
& \stackrel{(\ref{eq:dad_lrt_normal})}{=} %
\sum_{m=1}^{M} Q \left( \frac{\gamma}{\sqrt{n\sigma^2}} \right) \nonumber\\
&= M \cdot  Q \left( \frac{\gamma}{\sqrt{n\sigma^2}} \right).
\label{eq:pfa_dad_proof}
\end{align}
Since $\operatorname{p}(\mathbf{y}\mid\mathbf{x}_{\hat{m}}) \geq \operatorname{p}(\mathbf{y}\mid\mathbf{x}_{m})$, $P_\mathrm{MD}$ can be bounded by
\begin{align} 
P_\mathrm{MD} &\stackrel{(\ref{eq:dad_sub_optimal_JDD})}{=}%
\frac{1}{M} \sum_{m=1}^{M} \mathrm{Pr} \left\{\frac{
    \operatorname{p}(\mathbf{y}\mid\mathbf{x}_{\hat{m}})
    }
    {
    \operatorname{p}(\mathbf{y}\mid\mathbf{x}_{\O})
    }  < \gamma \;\middle |\; \mathbf{x}_m \right\} \nonumber \\
&\leq \frac{1}{M} \sum_{m=1}^{M} \mathrm{Pr} \left\{\frac{
    \operatorname{p}(\mathbf{y}\mid\mathbf{x}_{m})
    }
    {
    \operatorname{p}(\mathbf{y}\mid\mathbf{x}_{\O})
    } < \gamma \;\middle |\; \mathbf{x}_m \right\} \nonumber
    \\
&\stackrel{(\ref{eq:dad_lrt_normal})}{=} 1- Q \left( \frac{\gamma-n}{\sqrt{n\sigma^2}} \right). \label{eq:pmd_dad_proof}
\end{align}
Finally, an upper bound on the inclusive error rate can be obtained by again applying the \ac{UB}
\begin{align*}
    P_\mathrm{IE} &\stackrel{(\ref{eq:dad_sub_optimal_JDD})}{=} \frac{1}{M} \sum_{m=1}^{M} \mathrm{Pr} \left\{ \frac{
    \operatorname{p}(\mathbf{y}\mid\mathbf{x}_{\hat{m}})
    }
    {
    \operatorname{p}(\mathbf{y}\mid\mathbf{x}_{\O})
    } < \gamma ~ \lor ~ \hat{m} \neq m  \;\middle |\; \mathbf{x}_m \right\} \nonumber \\ 
    &\stackrel{\mathrm{UB}}{\leq} P_\mathrm{MD} + P_\mathrm{CW}. %
\end{align*}

\end{IEEEproof}
The bound proposed in Theorem \ref{thm:dad_ach_bound} is tight when the spacing of the codewords is large compared to the channel noise, i.e., for a high \ac{SNR} or a small code size $M$.
Then, the probabilitiy of a false alarm for each codeword is %
independent and the union bound from equation (\ref{eq:pfa_dad_proof}) is tight on $P_\mathrm{FA}$.
In this case, $\gamma$ can be chosen as
\begin{equation*}
    \gamma = \sqrt{n \sigma^2} \cdot Q^{-1} \left( \frac{\epsilon_\mathrm{FA}}{M} \right)
\end{equation*}
for a required $P_\mathrm{FA} \leq \epsilon_\mathrm{FA}$ in order to minimize the missed detection rate $P_\mathrm{MD}$ and inclusive error rate $P_\mathrm{IE}$.

Furthermore, if the probability of a decoding error is small, then the bound (\ref{eq:pmd_dad_proof}) is tight on $P_\mathrm{MD}$.

The bound can be transformed to the inequality below, allowing to find an achievable code size by iterative calculation:
\begin{equation*}
     M \leq \min  \left\{ \left\lfloor \frac{\epsilon_\mathrm{FA}}{Q \left( Q^{-1}(1-\epsilon_\mathrm{MD}) + \sqrt{\frac{n}{\sigma^2}} \right) } \right\rfloor, M^*(\tilde{p}_e,n,\sigma^2)
     \right\}
\end{equation*}
where $M^*$ is an achievable code size, and
\begin{equation*}
    \tilde{p}_e  = \epsilon_\mathrm{IE} - 1 + Q \left( Q^{-1} \left( \frac{\epsilon_\mathrm{FA}}{M} \right) - \sqrt{\frac{n}{\sigma^2}} \right).
\end{equation*}

\subsection{Preamble-Energy Detection}\label{ssec:hyped}
In this subsection, we propose a \ac{JDD} scheme for which the detecting complexity is reduced compared to \ac{DAD}. Starting from equation (\ref{eq:dad_sub_optimal_JDD}) and splitting the transmission into a preamble and codeword part, we get
\begin{align*}%
    \underbrace{\frac{
    \operatorname{p}(\mathbf{y}_\mathrm{p}|\mathbf{x}_\mathrm{p})
    }
    {
    \operatorname{p}(\mathbf{y}_\mathrm{p}|\O)
    }}_\text{Preamble}
    \cdot \underbrace{\frac{
    \max_{m'} \operatorname{p}(\mathbf{y}_\mathrm{c}|\mathbf{x}_{\mathrm{c},m'})
    }
    {
    \operatorname{p}(\mathbf{y}_\mathrm{c}|\O)
    }}_\text{Codeword}
    &\mathrel{\substack{\O\\\lessgtr\\\hat{m}}}
    \gamma,  \\
    \underbrace{\ln \left[\frac{
    \operatorname{p}(\mathbf{y}_\mathrm{p}|\mathbf{x}_\mathrm{p})
    }
    {
    \operatorname{p}(\mathbf{y}_\mathrm{p}|\O)
    } \right]}_{\triangleq \ell(\mathbf{y}_\mathrm{p})}
    + \underbrace{\ln \left[\frac{
    \max_{m'} \operatorname{p}(\mathbf{y}_\mathrm{c}|\mathbf{x}_{\mathrm{c},m'})
    }{
    \operatorname{p}(\mathbf{y}_\mathrm{c}|\O)
    }\right]}_{\triangleq \ell(\mathbf{y}_\mathrm{c})}
    &\mathrel{\substack{\O\\\lessgtr\\\hat{m}}}
    \ln \left[ \gamma  \right],
\end{align*} 
where the subscripts $\mathrm{p}$ and $\mathrm{c}$ indicate the preamble of length $n_\mathrm{p}$ and codeword of length $n_\mathrm{c}$, respectively. Applying the logarithm, we obtain the threshold test in terms of \acp{LLR}.
Since the preamble is known, we can rewrite for a \ac{BI-AWGN} channel $\ell(\mathbf{y}_\mathrm{p})=\sum_i\ell(y_{\mathrm{p},i})$.
Without loss of generality, we set the preamble $\mathbf{x}_\mathrm{p}=\mathbf{1}$ resulting in
\begin{equation*}
\ell (y_{\mathrm{p},i})=\frac{2y_{\mathrm{p},i}-1}{2\sigma^2}.
\end{equation*}
Evaluating $\ell(\mathbf{y}_\mathrm{c})$ exactly requires finding the \ac{ML}-codeword. 
Energy detection simplifies the computational complexity of the detector, as it assumes all bits to be \ac{i.i.d.}, which results in $\ell(\mathbf{y}_\mathrm{c})=\sum_i\ell(y_{\mathrm{c},i})$.
With the prior probabilities given as $\mathrm{Pr} \left\{x_{\mathrm{c},i} = +1 \right\} = p$ and $\mathrm{Pr} \left\{x_{\mathrm{c},i} = -1 \right\} = 1-p$, the \acp{LLR} for the codeword symbols can be calculated as in \cite{lancho2021jdd}%
\begin{align*}%
    \ell (y_{\mathrm{c},i}) 
    &= \ln{ \left[ p \exp \left( \frac{y_{\mathrm{c},i}}{\sigma^2} \right) + (1-p) \exp \left(- \frac{y_{\mathrm{c},i}}{\sigma^2} \right) \right] } - \frac{1}{2\sigma^2}.
\end{align*}
The total \ac{LLR} can then be computed as
\begin{align*}
    \ell (\mathbf{y}) 
    &= \sum_{i=0}^{n_{\mathrm{c}}-1}\ln \left[ p \exp \left( \frac{y_{\mathrm{c},i}}{\sigma^2} \right) + (1-p) \exp \left(- \frac{y_{\mathrm{c},i}}{\sigma^2} \right) \right] \\
    &+ \sum_{i=0}^{n_{\mathrm{p}}-1} \frac{y_{\mathrm{p},i}}{\sigma^2} - \frac{n}{2\sigma^2}.
\end{align*}
To simplify, we assume equal priors, i.e., $p=\frac{1}{2}$. This is motivated by the observation that skewing the input distribution reduces the decoding performance significantly, while the detection performance is only slightly improved. Then, we can simplify the \ac{LLR} to
\begin{equation*}%
    \ell (\mathbf{y}) = \sum_{i=0}^{n_{\mathrm{c}}-1} \ln \left[ \cosh \left( \frac{y_{\mathrm{c},i}}{\sigma^2} \right) \right] + \sum_{i=0}^{n_{\mathrm{p}}-1} \frac{y_{\mathrm{p},i}}{\sigma^2} - \frac{n}{2\sigma^2}.
\end{equation*}
With the obtained \ac{LLR}, the Neyman-Pearson decision rule can be applied to perform detection. The performance of the receiver is then
\begin{align*}
    P_\mathrm{FA} &= \mathrm{Pr} \left\{ \ell(\mathbf{y} ) \geq \gamma \mid \mathbf{x}_{\O} \right\}, \\
    P_\mathrm{MD} &= \mathrm{Pr} \left\{ \ell(\mathbf{y}) < \gamma \mid \mathbf{x}_{m} \right\}, \\
    P_\mathrm{IE} &= \mathrm{Pr} \left\{ \ell(\mathbf{y}) < \gamma ~ \lor ~ \hat{m} \neq m \mid \mathbf{x}_{m} \right\}.
\end{align*}
Both achievability and converse bounds can be calculated by Monte Carlo simulation for the false alarm and missed detection rate. The inclusive error rate is bounded by
\begin{equation}\label{eq:pie_bound_lower_upper}
    \max \left\{ P_\mathrm{MD}, P_\mathrm{CW} \right\} \leq P_\mathrm{IE} \leq P_\mathrm{MD} + P_\mathrm{CW}.
\end{equation}
For the decoding error rate $P_\mathrm{CW}$, finite-blocklength bounds for synchronous communication can be used, e.g. from \cite{polyanski2010bounds}.

A few remarks on the new \ac{HyPED} detection rule:
\begin{itemize}
    \item In order to reduce the computational complexity, the decision regions can be approximated by a function $f(\mathbf{y})$ that is chosen such that $ \left| f(\mathbf{y}) - \sum_i \ln [\cosh{(y_i)}] \right| $ is small on average.
    \item %
    The heuristic \ac{HyPED} rule \cite[Sec. 4B]{wang2011errorexponents} in equation (\ref{eq:wang_heuristic_jdd_peramble_energy}) is such an approximation, where $f(\mathbf{y})= \Vert \mathbf{y} \Vert$.
    \item The \Ac{HyPED} approach yields a better result than preamble based detection, as $\ell (\mathbf{y}_\mathrm{c})$ provides additional information about the transmitter state.
\end{itemize}

\section{Numerical Results}
In the following, we provide numerical results for the discussed detection schemes and their bounds in Fig.~\ref{fig:rate_over_blocklength_snrm3} and Fig.~\ref{fig:pie_over_snr}. In Subsection \ref{sub:bounds} we comment on the bounds in both plots, while Subsection \ref{sub:actual_performance} discusses the performance of actual systems.

\begin{figure}[htb]
\input{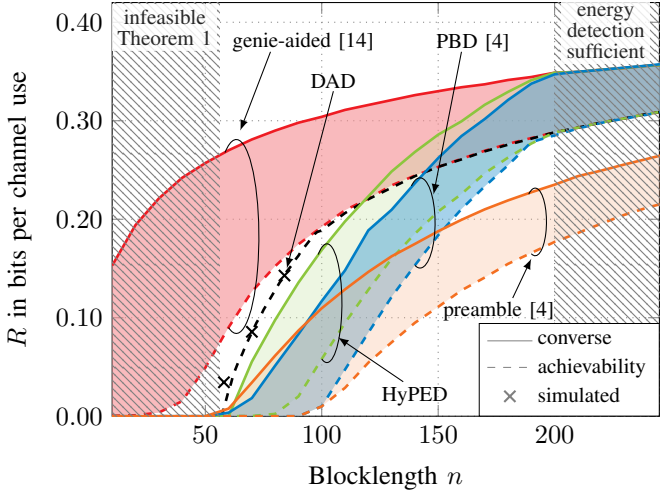}

\begin{tikzpicture}
\begin{axis}[
	width=\linewidth,
	height=.8\linewidth,
	grid style={dotted,gray},
	xmajorgrids,
	yminorticks=true,
	ymajorgrids,
    y tick label style={
        /pgf/number format/.cd,
            fixed,
            fixed zerofill,
            precision=2,
        /tikz/.cd
    },
	legend columns=1,
	legend style={at={(1,0)},anchor=south east, nodes=right, line width=0.1pt},
	legend cell align={left},
	xlabel={Blocklength $n$},
	ylabel={$R$ in bits per channel use},
	legend image post style={mark indices={}},
	mark size=1.5pt,
	line width = \lw,
	xmin=10,
	xmax=245,
	ymin=0,
	ymax=0.42
]
\addlegendimage{no marks, color=mittelgrau}
\addplot[color=rot, name path=genie_conv, line width = \lw, forget plot] table [x=n, y=genie_conv, col sep=comma] {lancho_snrm3.csv};\label{plt:genie_conv}
\addlegendentry{\footnotesize converse}
\addlegendimage{no marks, color=mittelgrau, dashed}
\addplot[color=rot, name path=genie_ach, dashed,forget plot,line width = \lw] table [x=n, y=genie_ach, col sep=comma] {lancho_snrm3.csv};\label{plt:genie_ach}
\addlegendentry{\footnotesize achievability}
\addplot[rot!50, fill opacity=0.6, forget plot] fill between[of=genie_conv and genie_ach];
\addlegendimage{only marks, mark=x, mark options={scale=2}, thick, color=mittelgrau};
\addlegendentry{\footnotesize simulated}

\addplot[color=black, dashed,line width = \lw] table [x=n, y=ach, col sep=comma]{decoder_aided_ach.csv};\label{plt:dad_ach}

\addplot[color=apfelgruen, name path=hyped_conv, forget plot,line width = \lw] table [x=n, y=conv, col sep=comma] {hyped.csv};\label{plt:hyped_conv}
\addplot[color=apfelgruen, name path=hyped_ach, dashed, forget plot,line width = \lw] table [x=n, y=ach, col sep=comma] {hyped.csv};\label{plt:hyped_ach}
\addplot[apfelgruen!50, fill opacity=0.3] fill between [of=hyped_conv and hyped_ach];

\addplot[color=mittelblau, name path=joint_conv,forget plot,line width = \lw] table [x=n, y=joint_conv, col sep=comma] {lancho_snrm3.csv};\label{plt:lancho_conv}
\addplot[color=mittelblau, name path=joint_ach, dashed,forget plot,line width = \lw] table [x=n, y=joint_ach, col sep=comma] {lancho_snrm3.csv}; \label{plt:lancho_ach}
\addplot[mittelblau!70, fill opacity=0.5] fill between[of=joint_conv and joint_ach];

\addplot[color=orange, name path=pre_conv,forget plot,line width = \lw] table [x=n, y=pre_conv, col sep=comma] {lancho_snrm3.csv};\label{plt:preamble_conv}
\addplot[color=orange, name path=pre_ach, dashed,forget plot,line width = \lw] table [x=n, y=pre_ach, col sep=comma] {lancho_snrm3.csv};\label{plt:preamble_ach}
\addplot[orange!50, fill opacity=0.3] fill between[of=pre_conv and pre_ach];

\draw[pattern=north west lines, pattern color=mittelgrau, draw=none] (axis cs: 0,0) rectangle (axis cs: 56,0.6);

\draw[pattern=north west lines, pattern color=mittelgrau, draw=none] (axis cs: 200,0) rectangle (axis cs: 250,0.6);

\addplot
[only marks, color=black, mark=x, mark options={scale=2}, thick]
table {
58  0.03448 
70  0.08571 
84  0.14286 
};\label{plt:system}

\node[fill=white, align=center, inner sep=2pt, opacity=0.85] at (axis cs:32.6, 0.395) {\footnotesize infeasible \\[-4pt] \footnotesize Theorem 1};
\node[fill=white, align=center, inner sep=2pt, opacity=0.85] at (axis cs:222, 0.39) {\footnotesize energy \\[-4pt] \footnotesize detection \\[-4pt] \footnotesize sufficient}; 

\draw (axis cs: 60, 0.09) arc(-110:120:0.3cm and 1.29cm);
\draw (axis cs: 100, 0.058125) arc(-110:125:0.2cm and 0.79cm);
\draw (axis cs: 140, 0.15322) arc(-110:125:0.2cm and 0.6cm);
\draw (axis cs: 190, 0.165) arc(-110:125:0.15cm and 0.45cm);

\draw[-latex] (axis cs:78,0.37) -- (axis cs: 65, 0.29);
\node[] at (axis cs: 92, 0.38){\footnotesize genie-aided \cite{polyanski2010bounds}};

\draw[-latex] (axis cs:105,0.33) -- (axis cs: 85, 0.15);
\node[] at (axis cs: 105, 0.34){\footnotesize \ac{DAD}};

\draw[-latex] (axis cs:135,0.035) -- (axis cs: 107, 0.08);
\node[] at (axis cs: 140, 0.02){\footnotesize \ac{HyPED}};

\draw[-latex] (axis cs:160,0.37) -- (axis cs: 148, 0.225);
\node[] at (axis cs: 165, 0.38){\footnotesize \ac{PBD} \cite{lancho2021jdd}};

\draw[-latex] (axis cs:182,0.12) -- (axis cs: 190, 0.16);
\node[] at (axis cs: 176, 0.107){\footnotesize preamble \cite{lancho2021jdd}};

\end{axis}
\end{tikzpicture}
\caption{Achievability (dashed) and converse (solid) bounds on the rate $R$ for $E_\mathrm{S}/N_0=-3\,\mathrm{dB}$, $\epsilon_\mathrm{FA}=10^{-4}$, $\epsilon_\mathrm{MD}=10^{-4}$, and $\epsilon_\mathrm{IE}=10^{-3}$ for different detection schemes. Performances of \ac{DAD} achieved in Monte Carlo simulation are marked by \ref{plt:system}.}
\label{fig:rate_over_blocklength_snrm3}
\end{figure}

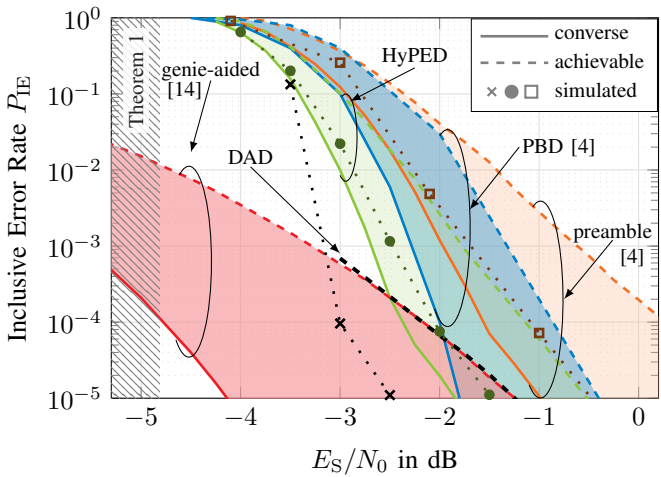
\begin{figure}[htb]
\begin{tikzpicture}

\begin{axis}[
	width=1\linewidth,
	height=.75\linewidth,
	xmajorgrids,
	yminorticks=true,
	ymajorgrids,
    yminorgrids,
    major grid style={hellgrau},
    minor grid style={dotted, hellgrau},
	legend columns=1,
	legend pos=north east,   
	legend cell align={left},
	ylabel={Inclusive Error Rate $P_\mathrm{IE}$},
	xlabel={$E_\mathrm{S}/N_0$ in dB},
	legend image post style={mark indices={}, scale=1.5},
	line width=\lw,
	legend style={at={(1.,1.)},anchor=north east, line width=0.1pt, cells={line width=0.8pt, mark size=1.5pt}},
	ymode=log,
	mark size=1.5,
	xmin=-5.3,
	xmax=0.2,
	ymin=1e-5,
	ymax=1,
	combo legend/.style={
	    legend image code/.code={
            \draw [style={mittelgrau, only marks, mark=x, mark size=1.5pt}] plot coordinates {(1mm,0cm)};
            \draw [style={mittelgrau, mark=*, mark size= 1.5pt, only marks}] plot coordinates {(2.75mm,0cm)};
            \draw [style={mittelgrau, only marks, mark=square, mark size= 1.5pt}] plot coordinates {(4.5mm,0cm)};
	    }
	}
]

\addlegendimage{no marks, color=mittelgrau, line width=\lw}
\addplot[color=orange, line width=\lw, mark size=1.5, mark options={solid}, name path=pre_conv, forget plot] 
table [x=snr, y=conv, col sep=comma]{pie_preamble.csv};
\addlegendentry{\footnotesize converse}

\addlegendimage{no marks, dashed, color=mittelgrau, line width=\lw}
\addplot[color=orange, line width=\lw, mark size=1.5, mark options={solid}, name path=pre_ach, dashed, forget plot] 
table [x=snr, y=ach, col sep=comma]{pie_preamble.csv};
\addlegendentry{\footnotesize achievable}
\addplot[orange!50, fill opacity=0.3, forget plot] fill between[of=pre_conv and pre_ach];

\addplot[color=mittelblau, line width=\lw, mark size=1.5, mark options={solid}, dashed, name path=lancho_ach, forget plot] 
table [x=snr, y=ach, col sep=comma]{pie_lancho.csv};
\addplot[color=mittelblau, line width=\lw, mark size=1.5, mark options={solid}, name path=lancho_conv, forget plot] 
table [x=snr, y=conv, col sep=comma]{pie_lancho.csv};
\addplot[mittelblau!70, fill opacity=0.5, forget plot] fill between[of=lancho_conv and lancho_ach];

\addplot[color=apfelgruen, line width=\lw, mark size=1, mark options={solid}, name path=hyped_ach, dashed, forget plot]
table [x=snr, y=ach, col sep=comma]{pie_hyped.csv};
\addplot[color=apfelgruen, line width=\lw, mark size=1, mark options={solid}, name path=hyped_conv, forget plot]
table [x=snr, y=conv, col sep=comma]{pie_hyped.csv};
\addplot[apfelgruen!50, fill opacity=0.3, forget plot] fill between[of=hyped_conv and hyped_ach];

\addplot[color=black, line width=1.5pt, mark size=1, mark options={solid}, dashed, forget plot]
table [x=snr, y=ach, col sep=comma]{pie_dad.csv};
\addplot[color=rot, line width=\lw, mark size=1, mark options={solid}, dashed, name path=genie_ach, forget plot, forget plot]
table[x=snr, y=ach, col sep=comma]{pie_genie.csv};
\addplot[color=rot, line width=\lw, mark size=1, mark options={solid}, name path=genie_conv, forget plot, forget plot]
table [x=snr, y=conv, col sep=comma]{pie_dad.csv};
\addplot[rot!50, fill opacity=0.6, forget plot] fill between[of=genie_conv and genie_ach];

\addplot[color=black, mark=x, line width=\lw, mark size=2.5, mark options={solid}, loosely dotted, forget plot] table [col sep=comma]
{
-3.5,0.13353
-3., 9.6e-5
-2.5, 1.1e-5
};\label{plt:pie_dad}

\addplot[color=apfelgruen!50!black, mark=*, mark size= 1.5pt, line width=\lw, loosely dotted, mark options={solid}, forget plot] table [col sep=comma]
{
-4, 0.6514
-3.5, 0.2015
-3, 0.02213
-2.5, 0.001157
-2., 7.54e-5
-1.5, 1.11e-5
};\label{plt:pie_hyped}

\addplot[color=orange!50!black, mark=square, mark size= 1.5pt, line width=\lw, loosely dotted, mark options={solid}, forget plot] table [col sep=comma]
{
-4.1, 0.9110
-3, 0.25752
-2.1, 0.004849857
-1, 7.20279864853152e-05
0, 1.688468929674371e-06
};\label{plt:pie_preamble}

\draw[pattern=north west lines, pattern color=mittelgrau, draw=none] (axis cs: -5.3,1e-5) rectangle (axis cs: -4.82,1);

\node[fill=white, align=center, inner sep=2pt, opacity=0.85, rotate=90] at (axis cs:-5.05, 1.5e-1) {\footnotesize Theorem 1};

\draw (axis cs: -4.65, 5.85e-5) arc(-125:115:0.3cm and 1.27cm);
\draw (axis cs: -3, 8.5e-3) arc(-120:110:0.15cm and 0.58cm);
\draw (axis cs: -1.1, 1.5e-5) arc(-120:110:0.3cm and 1.31cm);
\draw (axis cs: -2.05, 1.5e-4) arc(-125:110:0.3cm and 1.31cm);

\draw[-latex] (axis cs:-4.3,1.5e-1) -- (axis cs: -4.5, 1.7e-2);
\node[] at (axis cs: -4.3,2e-1){\footnotesize genie-aided};
\node[]at (axis cs: -4.6,1.1e-1){\footnotesize \cite{polyanski2010bounds}};

\draw[-latex] (axis cs:-3.9,1.1e-2) -- (axis cs: -3., 8e-4);
\node[] at (axis cs: -3.9,1.5e-2){\footnotesize \ac{DAD}};

\draw[-latex] (axis cs:-2.25,2.2e-1) -- (axis cs: -2.9, 8e-2);
\node[] at (axis cs: -2.25, 3e-1){\footnotesize \ac{HyPED}};

\draw[-latex] (axis cs:-1.2,2e-2) -- (axis cs: -1.7, 3e-3);
\node[] at (axis cs: -0.8,2e-2){\footnotesize \ac{PBD}\cite{lancho2021jdd}};

\draw[-latex] (axis cs:-0.26,1.2e-3) -- (axis cs: -0.75, 2e-4);
\node[] at (axis cs: -0.26,1.5e-3){\footnotesize preamble};
\node[] at (axis cs: -0.1,7e-4){\footnotesize \cite{lancho2021jdd}};

\addlegendimage{combo legend}
\addlegendentry{\footnotesize simulated};

\end{axis}
\end{tikzpicture}
\caption{$P_\mathrm{IE}$ over \ac{SNR} of achievability (dashed) and converse (solid) bounds and simulated systems (dotted) for $\epsilon_\mathrm{FA}=\epsilon_\mathrm{MD}= 10^{-4}$ and
blocklength $n=84$ and $k=12$. The simulated plots are marked by \ref{plt:pie_dad} for \ac{DAD}, \ref{plt:pie_hyped} 
for \ac{HyPED} and \ref{plt:pie_preamble} for preamble-based detection.}
\label{fig:pie_over_snr}
\end{figure}

\subsection{Bounds}\label{sub:bounds}

In Fig.~\ref{fig:rate_over_blocklength_snrm3}, we show the converse and achievability bounds on the information rate for the previously discussed detection schemes with respect to the blocklength $n$. 
The \ac{SNR} is set to $-3\,\mathrm{dB}$ \footnote{Due to different definitions of the \ac{SNR}, this is equivalent to $0\,\mathrm{dB}$ in \cite{lancho2021jdd}.} and requirements on the error rates are $\epsilon_\mathrm{FA}=10^{-4}$, $\epsilon_\mathrm{MD}=10^{-4}$ and $\epsilon_\mathrm{IE}=10^{-3}$.
For each blocklength, the solid lines show converse bounds on the information rate, while the dashed lines show achievability bounds.
For reference, the preamble-based detection, \ac{PBD} and genie aided curves from \cite{lancho2021jdd} are also shown.
All achievability results use the \ac{DT} bound, and all converse results use the meta-converse bound, both from \cite{polyanski2010bounds}.
The hatched region on the left labeled ``infeasible'' is the result from Theorem \ref{thm:converse_blocklength}. Thus, no communication system with blocklength $n \le 56$ can exist while meeting the requirements on the error rates.
The hatched region on the right marks the regime where all \ac{JDD} schemes considered approach the genie-aided curves. %
Thus, in this region, the decoding problem dominates, while the detection problem can be solved with any distribution based scheme.
The \ac{HyPED} bounds for the detection problem ($P_\mathrm{FA}$ and $P_\mathrm{MD}$) are computed via Monte-Carlo simulation. Subsequently, $P_\mathrm{IE}$ can be bounded by (\ref{eq:pie_bound_lower_upper}).
In comparison to \ac{PBD}, the \ac{HyPED} achievability and converse bound %
evaluate to larger information rates.
The achievability bound for the \ac{DAD} system (Theorem \ref{thm:dad_ach_bound}) outperforms all other proposed schemes.
For the considered parameters and small blocklengths ($n\geq59$), the bound guarantees the existence of systems close to the converse (Theorem \ref{thm:converse_blocklength}).
Further, the achievability bound approaches the genie-aided achievability bound at $n=100$,
indicating near optimal detection performance for $100$ channel uses less than the other \ac{JDD} schemes.

Fig.~\ref{fig:pie_over_snr} shows the inclusive error rate $P_\mathrm{IE}$ versus the \ac{SNR} for $n=84$ channel uses and $k=12$ information bits per block with requirements on the false alarm and missed detection rates $\epsilon_\mathrm{FA}=10^{-4}$ and $\epsilon_\mathrm{MD}=10^{-4}$. Again, solid and dashed lines show converse and achievability bounds, respectively.
Similar to Fig.~\ref{fig:rate_over_blocklength_snrm3}, the hatched region on the left results from Theorem \ref{thm:converse_blocklength}.
Thus, for \acp{SNR} below $-4.8\,\mathrm{dB}$, no communication system can exist that fulfills the requirements.
The curves are only plotted, if the detection performance meets the error-rate constraints, thus, beginning only for the \ac{SNR} above some threshold.
Overall, the \ac{JDD} schemes indicate possible gains over preamble based detection, especially for increasing \ac{SNR}.
The \ac{DAD} achievability bound only exists for \acp{SNR} $E_\mathrm{S}/N_0\geq-3\,\mathrm{dB}$. It matches the genie-aided achievability bound, and, therefore, tightly approximates the optimal detector and decoder.
Similar to the behavior for large blocklengths in Fig.~\ref{fig:rate_over_blocklength_snrm3}, for increasing \acp{SNR} the difference between the \ac{HyPED} and the \ac{PBD} bound decreases.

\subsection{Performance of Actual Systems} \label{sub:actual_performance}
To give an intuition of how actual systems perform in relation to the bounds, we construct practical preamble-based and \ac{JDD} systems for the discussed detection rules and obtain their operating points using Monte Carlo simulation. The decoding is performed using the \ac{ML} decision rule.

Note that for \ac{DAD}, the code influences not only the decoding but also the detecting performance. In particular, optimizing the code for a low decoding error rate spaces all codewords as distant from each other as possible, while optimizing with respect to a good detection performance yields a code with closely clustered codewords. Thus, a trade-off has to be made when designing the code used for \ac{JDD}.
However, designing such optimized codes is out of scope of this paper and, hence, we employ codes optimized for the decoding problem only, i.e., solely maximizing their minimum distance.

For Fig.~\ref{fig:rate_over_blocklength_snrm3}, the codes for the \ac{DAD} scheme have the parameters $(58,2)$, $(70,6)$ and $(84, 12)$.
The simulated systems lie close to the achievability bound from Theorem \ref{thm:dad_ach_bound}, and beyond the converse bounds for the other non-genie-aided schemes.
Thus, the simulated \ac{DAD} systems achieve higher information rates than the other considered schemes allow for.

The codes used for Fig.~\ref{fig:pie_over_snr} are the best known  $(n_\mathrm{c},k=12)$ binary linear codes in terms of minimum distance taken from \cite{codetables}.
For \ac{DAD}, $n_\mathrm{c}=84$ for \acp{SNR} $-3\,\mathrm{dB}$ and above, while for \ac{HyPED}, preamble-based detection and \ac{DAD} with \acp{SNR} below $-3\,\mathrm{dB}$, a preamble of length $n_\mathrm{p}$ is prepended with $n_\mathrm{p}+n_\mathrm{c}=84$. The length of the preamble is optimized for each operating point.
For $P_\mathrm{IE}=10^{-4}$, \ac{DAD} achieves a gain of $1\,\mathrm{dB}$ over \ac{HyPED}, which in turn gains $1\,\mathrm{dB}$ over preamble detection.
For very low \acp{SNR}, the \ac{DAD} system does not show a significant gain over \ac{HyPED} due to the high decoding error rate.

\section{Conclusion and Outlook}
For a slotted asynchronous system with transmission over the \ac{BI-AWGN} channel, we have shown new bounds for the \ac{JDD} problem in the finite blocklength regime.
Furthermore, we demonstrated that \ac{JDD} systems using \ac{DAD} or \ac{HyPED} outperform preamble-based detection and yield bounds suggesting higher rates than \ac{PBD} with \ac{i.i.d.} priors. Especially \ac{DAD} performs well, approaching the achievable information rate for synchronous transmission quickly.

Remaining open problems are finding short-blocklength bounds for optimal detection and decoding and a converse bound on the information rate of \ac{DAD}.

\bibliographystyle{IEEEtran}
\bibliography{references}
\end{NoHyper}

\end{document}